\documentclass[twocolumn,pra, showpacs,superscriptaddress]{revtex4}
\usepackage{epsfig,graphicx,times}
\usepackage{amstext}
\usepackage{amsmath}
\usepackage{amssymb}
\usepackage{graphicx}
\usepackage{latexsym}
\usepackage{bm}

\begin{document}

\title{Correlated Photons and Collective Excitations of Cyclic Atomic
Ensemble}
\author{Yong Li}
\affiliation{Institute of Theoretical Physics \& Interdisciplinary Center of Theoretical
Studies, Chinese Academy of Sciences, Beijing, 100080, China}
\author{Li Zheng}
\affiliation{College of Information Science and Engineering, Dalian Institute of Light
Industry, Dalian, 116034, China}
\author{Yu-xi Liu}
\affiliation{Frontier Research System, The Institute of Physical and Chemical Research
(RIKEN), Wako-shi 351-0198, Japan}
\author{C. P. Sun}
\email{suncp@itp.ac.cn} \homepage{http://www.itp.ac.cn/~suncp}
\affiliation{Institute of Theoretical Physics \& Interdisciplinary
Center of Theoretical Studies, Chinese Academy of Sciences, Beijing,
100080, China}
\date{\today}

\begin{abstract}
We systematically study the interaction between two quantized
optical fields and an cyclic atomic ensemble driven by a classic
optical field. This so-called atomic cyclic ensemble consists of the
three-level atoms with $\Delta$-type transitions due to the symmetry
breaking, which can also be implemented in the superconducting
quantum  circuit [Phys. Rev. Lett. \textbf{95}, 087001 (2005)]. We
explore the dynamic mechanisms to creating  the quantum
entanglements among  photon states, and between  photons and atomic
collective excitations by the coherent manipulation of the
atom-photon system. It is shown that the quantum information can be
completely transferred from one quantized optical mode to another,
and the quantum information carried by the two quantized optical
fields can be stored in the collective modes of this atomic ensemble
by adiabatically controlling the classic field Rabi frequencies.
\end{abstract}

\maketitle

\section{Introduction}

The electric-dipole selection rule does not allow one- and two-photon
processes to coexist for given initial and final states in quantum
systems with a center of inversion symmetry, where all states have
well-defined parities~\cite{group}. Because one-photon transitions,
resulted from the electric-dipole interaction between two non-degenerate
states, require that these states have opposite parities; but two-photon
process needs those states to have the same parity~\cite{group}. However,
the one- and two-photon processes can coexist in systems with lack of
inversion symmetry,
e.g., in the semiconductor systems \cite%
{Kurizki89,Dupont95,Atanasov96,kral99}. Then, the magnitude and direction of
the photocurrent in these systems \cite{Kurizki89,Dupont95,Atanasov96,kral99}
can be controlled by using two different optical paths.

Ref.~\cite{Kral0103} showed that electric-dipole transition between any two
states is allowed for chiral molecules and their mirror images due to lack
of inverse symmetry. It means that the one- and two-photon processes can
also coexist in these systems. Then, the same initial and final states can
be connected by two different paths, which result in an interference effect
for final state. The different relative phase differences for pulses of the
two optical pathways will result in different interference fringes. This
implies that final state can be controlled by choosing applied pulse phases.
Using an example in Ref.~\cite{Kral0103}, the coherent population transfer
(CPT) was studied in a three-level system with cyclic transitions, induced
by three classical fields. Different from the usual $\Lambda $-type atoms~%
\cite{Bergmann}, the CPT in these systems is controlled not only by the
amplitudes of the electric-dipole transition elements, but also by the
phases of applied pulses.

Recently, the microwave control of the quantum states has been investigated
for superconducting quantum circuits~\cite{liuyx-prl}, called ``artificial
atoms", which possess discrete energy levels. The optical selection rule of
microwave-assisted transitions is carefully analyzed for this artificial
atom. It was shown~\cite{liuyx-prl} that the electric-dipole like transition
can appear for any two different states when the symmetry of the potential
of the artificial atom is broken by changing microwave bias. Then, so-called
$\Delta $-type or cyclic transitions can be formed for the lowest three
levels. And the populations of these states can be selectively transferred
by adiabatically controlling both the amplitudes and phases of the applied
microwave pulses.

The previous investigations, e.g., in Refs. \cite%
{Kurizki89,Dupont95,Atanasov96,kral99,Kral0103,liuyx-prl}, only focus on a
single three-level system with $\Delta $-type or cyclic transitions, induced
by the three classical fields. In contrast to the above examples~\cite%
{Kurizki89,Dupont95,Atanasov96,kral99,Kral0103,liuyx-prl}, the
electric-dipole interaction cannot induce cyclic transitions among three
energy levels for a usual atom, due to its symmetry and well-defined
parities of its eigenstates. To have cyclic or $\Delta $-type transitions, a
coherent radio-frequency field is required to apply such that it can induce
a magnetic dipole (or an electric quadrupole) transition~\cite%
{Buckle,Kosachiov,Fleischhauer} between two levels, e.g., two lower (or
higher) levels which are forbidden to the electric-dipole transition in the $%
\Lambda $ (or $V$)-type atoms~\cite{Bergmann}.

In this paper, we will investigate the collective effects of photonic
emissions and excitations of a cyclic three-level system (such as atomic
ensemble with symmetry broken, or the chiral molecular gases \cite{Kral0103}%
, or manual \textquotedblleft atomic" array with symmetry broken~\cite%
{liuyx-prl,Siewert}) where the quantum transitions is induced by two
quantized fields and a classical one. We will focus on the photonic
properties of emissions resulting from such cyclic transitions, such as the
two-mode photon entanglement, the quantum state exchange and swapping
between the two-mode optical field and the two collective modes of atomic
excitations.

In more details, by utilizing the collective operator approach and the
hidden dynamic symmetry as recently discovered~\cite{sun-prl} for the
three-level $\Lambda $-type atomic ensemble coupled to a classical control
field and a quantum probe field, both the adiabatic and dynamic properties
for the system of the photons and atomic ensemble are studied
systematically. Different from the case of three-level $\Lambda $ (or $V$,
or $\Xi $)-type atomic ensemble, due to the interference between one- and
two-photon processes in the system of $\Delta $-type atomic ensemble, we
find that the electromagnetically induced transparency (EIT) phenomenon~\cite%
{EIT}, appeared in $\Lambda $-type system, does not exist here. In stead of
dark-state polariton operators~\cite{sun-prl}, a general set of polariton
operators is introduced to describe the collective motions of the whole
system when the excitation to high energy levels is low. Some novel results
are obtained. For example, the entanglement between two quantum optical
fields is tunable via classical field, applied to the $\Delta $-type atomic
ensemble.

Our paper is organized as follows. In Sec.~II, the model is described and
the polariton operators are introduced in the limit of the low excitation.
In Sec.~III, the entanglement between, e.g., the atomic ensemble and
quantized fields, or two different optical modes, is discussed. In Sec.~IV,
we analyze the information transfer from the quantized fields to the atomic
ensemble by adiabatic passage, and study the storage of photon information
via atomic ensemble.

\section{Atomic ensemble with cyclic transitions and polariton excitations}

\begin{figure}[tbp]
\includegraphics[width=6cm,height=6cm]{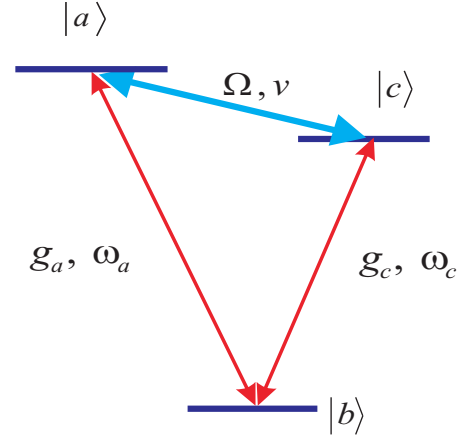}
\caption{(Color online) Three-level cyclic atoms are resonantly coupled
to two quantized fields and a classical field via electric-dipole
interaction.} \label{fig1}
\end{figure}

We consider an ensemble with $N$ identical three-level \textquotedblleft
atoms" (such as atomic ensemble or the chiral molecular gases \cite{Kral0103}
or manual \textquotedblleft atomic" array with symmetry broken~\cite%
{liuyx-prl}) interacting with electromagnetic fields. Each atom has cyclic
or $\Delta $-type transitions shown in Fig.~\ref{fig1}, where a lower level $%
|b\rangle $ is coupled to two higher levels $|a\rangle $ and $|c\rangle $ by
quantized fields through the electric-dipole interaction; and two higher
states are coupled by a classical field with a frequency $\nu $ through the
electric-dipole interaction. The Hamiltonian of the interacting system is
given as ($\hbar =1$)
\begin{eqnarray}
H_{ori} &=&\omega _{a}a^{\dagger }a+\omega _{b}b^{\dagger }b+\omega
_{ab}\sum_{j=1}^{N}\sigma _{\mathrm{aa}}^{(j)}+\omega
_{cb}\sum_{j=1}^{N}\sigma _{\mathrm{cc}}^{(j)}  \notag  \label{hamil01} \\
&&+g_{a}\sum\limits_{j}\mathrm{e}^{i\mathbf{K}_{a}\cdot \mathbf{r}%
_{j}}\sigma _{\mathrm{ab}}^{(j)}a+g_{b}\sum\limits_{j}\mathrm{e}^{i\mathbf{K}%
_{b}\cdot \mathbf{r}_{j}}\sigma _{\mathrm{cb}}^{(j)}b \\
&&+\Omega ^{\prime }\mathrm{e}^{-i\omega _{\nu }t}\sum_{j}\mathrm{e}^{i%
\mathbf{K}_{v}\cdot \mathbf{r}_{j}}\sigma _{\mathrm{ac}}^{(j)}+\text{h.c.}.
\notag
\end{eqnarray}%
Here, $\sigma _{\mathrm{mn}}^{(j)}=|m\rangle _{jj}\langle n|$, with $%
m,n=a,b,c$ but $n\neq m$, are the flip operators between the levels $%
|m\rangle _{j}$ and $|n\rangle _{j}$ of the $j$-th atom. $\sigma _{\mathrm{mm%
}}^{(j)}=|m\rangle _{jj}\langle m|$ ($m=a,c$) represent the population
operators. $a$ ($a^{\dagger }$) and $b$ ($b^{\dagger }$) are the
annihilation (creation) operators of the two quantized light fields, with
the angular frequencies (wave vectors) $\omega _{a}$ ($\mathbf{K}_{a}$) and $%
\omega _{b}$ ($\mathbf{K}_{b}$), respectively. The parameters
$g_{a}$ and $ g_{b}$ denote the coupling constants associated with
two quantized fields, coupling to the atomic transitions
$|a\rangle \rightarrow |b\rangle $ and $|c\rangle \rightarrow
|b\rangle $ respectively. Here, we assume that coupling constants
$g_{a}$ and $g_{b}$ of all atoms to the two quantized fields are
identical. $\omega _{ab}$ and $\omega _{cb}$ are the angular
frequencies of the atomic transitions $|a\rangle \rightarrow
|b\rangle $ and $|c\rangle \rightarrow |b\rangle $, respectively.
$\Omega ^{\prime }$ is the Rabi frequency related to the atomic
transition $|a\rangle \rightarrow |c\rangle $, driven by the
classic
field with the angular frequency $\omega _{\nu }$ and the wave vector $%
\mathbf{K}_{\nu }$.

For the sake of simplicity, we assume that three light fields are \textit{%
resonantly} coupled to the relevant atomic transitions, that is, $\omega
_{ab}=\omega _{a}$, $\omega _{cb}=\omega _{b}$, and $\omega _{ac}=\omega
_{\nu }$. Thus, in the interaction picture, the Hamiltonian~(\ref{hamil01})
can be simplified to
\begin{eqnarray}
H &=&g_{a}a\sum\limits_{j}\mathrm{e}^{i\mathbf{K}_{a}\cdot \mathbf{r}%
_{j}}\sigma _{\mathrm{ab}}^{(j)}+g_{b}b\sum\limits_{j}\mathrm{e}^{i\mathbf{K}%
_{b}\cdot \mathbf{r}_{j}}\sigma _{\mathrm{cb}}^{(j)}  \notag \\
&&+\Omega ^{\prime }\sum_{j=1}^{N}\mathrm{e}^{i\mathbf{K}_{\nu }\cdot
\mathbf{r}_{j}}\sigma _{\mathrm{ac}}^{(j)}+\text{h.c.}.  \label{hamil02}
\end{eqnarray}
Considering that any quantum state is allowed to have a global constant
difference of phase factor, one can redefine new atomic states with the
phase factors \cite{scully2} as
\begin{equation}
\left\vert a^{\prime }\right\rangle _{j}=\mathrm{e}^{i\mathbf{K}_{a}\cdot
\mathbf{r}_{j}}\left\vert a\right\rangle _{j},\ \left\vert b^{\prime
}\right\rangle _{j}=\left\vert b\right\rangle _{j},\left\vert c^{\prime
}\right\rangle _{j}=\mathrm{e}^{i\mathbf{K}_{b}\cdot \mathbf{r}%
_{j}}\left\vert c\right\rangle _{j}.
\end{equation}%
We further assume that the \textit{momenta} $K_{a}$, $K_{b}$, and $K_{\nu }$
for the three light fields satisfy the \textit{conversation} condition
\begin{equation}
\mathbf{K}_{b}+\mathbf{K}_{\nu }-\mathbf{K}_{a}=0.
\end{equation}%
Then, after the factors $\exp ({i\mathbf{K}_{a}\cdot \mathbf{r}_{j})}$, $%
\exp ({i\mathbf{K}_{b}\cdot \mathbf{r}_{j})}$, and $\exp ({i\mathbf{K}%
_{v}\cdot \mathbf{r}_{j})}$ are absorbed into atomic states, the
Hamiltonian~(\ref{hamil02}) can be rewritten as
\begin{equation}
H=g_{a}a\sum\limits_{j}\sigma _{\mathrm{ab}}^{(j)}+g_{b}b\sum\limits_{j}%
\sigma _{\mathrm{cb}}^{(j)}+\Omega ^{\prime }\sum_{j}\sigma _{\mathrm{ac}%
}^{(j)}+h.c..  \label{hamiltonian}
\end{equation}%
Here, we still use the notations $\left\vert m\right\rangle $ and $\sigma _{%
\mathrm{mn}}^{(j)}$ to denote $\left\vert m^{\prime }\right\rangle $ and $%
\sigma _{\mathrm{m^{\prime }n^{\prime }}}^{(j)}$ ($m^{\prime },\,n^{\prime
}=a^{\prime },\,b^{\prime },\,c^{\prime }$).

Obviously, the above Hamiltonian describes the homogeneous couplings of the
three light fields to atoms in the ensemble. This homogeneity means that
there exist various collective excitations that can be characterized by the
following operators~\cite{sun-prl}
\begin{eqnarray}
T^{-} &=&\sum_{j=1}^{N}\sigma _{\mathrm{ca}}^{(j)},\text{ \ \ }%
T^{z}=\sum_{j=1}^{N}\left( \sigma _{\mathrm{aa}}^{(j)}-\sigma _{\mathrm{cc}%
}^{(j)}\right) ,  \notag \\
\ A &=&\frac{1}{\sqrt{N}}\sum_{j=1}^{N}\sigma _{\mathrm{ba}}^{(j)},\text{ }C=%
\frac{1}{\sqrt{N}}\sum_{j=1}^{N}\sigma _{\mathrm{bc}}^{(j)}.
\end{eqnarray}%
Therefore, by using the above collective operators, the Hamiltonian~(\ref%
{hamiltonian}) can be expressed as
\begin{equation}
H=g_{a}\sqrt{N}a\,A^{\dagger }+g_{b}\sqrt{N}b\,C^{\dagger }+\Omega ^{\prime
}T^{+}+\text{h.c.}.  \label{h2}
\end{equation}%
Eq.~(\ref{h2}) implies that there exists a dynamic symmetry in the
considered system. This symmetry is characterized by Lie algebra generators $%
A$, $C$, $T^{-}$ and their complex conjugates $A^{\dagger }$, $C^{\dagger }$%
, $T^{+}$ (also $T^{z}$), that satisfy the following commutation relations
\begin{eqnarray}
\lbrack A,\,C^{\dagger }] &=&[A,\,C]\,=\,0,  \notag \\
\lbrack A,\,A^{\dagger }] &=&1,\,\,\,\,\,\,[C,\,C^{\dagger }]\,=\,1,  \notag
\\
\lbrack T^{-},\,C^{\dagger }] &=&0,\,\,\,\,\,\,[T^{+},\,C^{\dagger
}]\,=\,A^{\dagger }, \\
\lbrack T^{-},\,A^{\dagger }] &=&C^{\dagger },\,\,\,[T^{+},\,A^{\dagger
}]\,=\,0,  \notag \\
\lbrack T^{+},T^{-}] &=&T^{z}  \notag
\end{eqnarray}%
in the large $N$ and low excitation
limit~\cite{sun-prl,Jin03,Li04}. Where the low excitation means
that the most atoms are in the ground state,  only a few of them
are excited into the higher states. In this case, the average
numbers $\left\langle A^{\dagger }A\right\rangle$ and
$\left\langle C^{\dagger }C\right\rangle$ of the atoms in the two
excited states satisfy the condition $\left\langle A^{\dagger
}A\right\rangle /N<<1$ and $\left\langle C^{\dagger
}C\right\rangle /N<<1$. It means that two independent bosonic
modes ($A$ and $C$) of the atomic collective excitation exist in
the ensemble.

Since the complex coupling constants can be rewritten as $g_{a}=g_{a}^{0}%
\mathrm{\exp }[i\varphi _{a}]$, $g_{b}=g_{b}^{0}\mathrm{\exp }[i\varphi
_{c}] $, and $\Omega ^{\prime }=\Omega \mathrm{\exp }[i\varphi _{v}]$, where
$g_{a}^{0}$ and $g_{b}^{0}$ are positive real numbers, however $\Omega $ is
a real number. Then the phases $\varphi _{a}$ and $\varphi _{b}$ can be
absorbed into the operators $A$ and $C$ as follows: $A^{\dagger }\mathrm{%
\exp }[i\varphi _{a}]\rightarrow A^{\dagger }$, $C^{\dagger }\mathrm{\exp }%
[i\varphi _{c}]\rightarrow C^{\dagger }$. In this case, the operator $T^{+}$
should be changed as: $T^{+}\rightarrow T^{+}\mathrm{\exp }[i(\varphi
_{c}-\varphi _{a})]$. Without loss of generality, we now consider a simple
case with $g_{a}^{0}\sqrt{N}=$ $g_{b}^{0}\sqrt{N}\equiv g_{N}$. Under these
conditions, the Hamiltonian~(\ref{h2}) can be represented as
\begin{equation}
H=g_{N}aA^{\dagger }+g_{N}bC^{\dagger }+\Omega \mathrm{e}^{i\varphi }T^{+}+%
\text{h.c.},  \label{hh2}
\end{equation}%
where $\varphi =\varphi _{v}+\varphi _{c}-\varphi _{a}$. The transform from
Eq.~(\ref{h2}) to Eq.~(\ref{hh2}) means that only the total phase of the
three Rabi frequencies ($g_{a}$, $g_{b}$, and $\Omega ^{\prime }$) is
involved in the dynamical evolution.

The Hamiltonian~(\ref{hh2}) can be diagonalized by using polariton operators
$D_{i}$ ($i=1,\,2,\,3,\,4$) as
\begin{equation}
H=\sum_{i=1}^{4}\varepsilon _{i}D_{i}^{\dagger }D_{i},  \label{h3}
\end{equation}%
here the polariton operators
\begin{eqnarray}
D_{1,2} &=&\frac{\sin \theta }{\sqrt{2}}(a\pm b\mathrm{e}^{i\varphi })+\frac{%
\cos \theta }{\sqrt{2}}(C\mathrm{e}^{i\varphi }\pm A),  \label{pola12} \\
D_{3,4} &=&\frac{\cos \theta }{\sqrt{2}}(a\pm b\mathrm{e}^{i\varphi })-\frac{%
\sin \theta }{\sqrt{2}}(C\mathrm{e}^{i\varphi }\pm A),  \label{pola34}
\end{eqnarray}%
describe the normal bosonic modes with frequencies
\begin{eqnarray}
\varepsilon _{1} &\equiv &-\varepsilon _{2}=\frac{\Omega +\sqrt{\Omega
^{2}+4g_{N}^{2}}}{2}, \\
\text{ }\varepsilon _{3} &\equiv &-\varepsilon _{4}=\frac{\Omega -\sqrt{%
\Omega ^{2}+4g_{N}^{2}}}{2}.
\end{eqnarray}%
In Eqs.~(\ref{pola12}) and ~(\ref{pola34}), the first indexes of the l.h.s
correspond to the above symbols of the r.h.s, and
\begin{equation}
\theta =\arctan \frac{2g_{N}}{\Omega +\sqrt{\Omega ^{2}+4g_{N}^{2}}}.
\label{eq:16}
\end{equation}%
It is obvious $\theta \in \lbrack 0,\pi /2]$ for the positive real numbers $%
g_{a}^{0}$, $g_{b}^{0}$, and real number $\Omega $. From Eq.~(\ref{h3}), the
eigenstates of the system can be given as
\begin{eqnarray}
\left\vert \Psi _{lmnk}\right\rangle &=&\left\vert l,m,n,k\right\rangle
_{D_{1}D_{2}D_{3}D_{4}}  \notag \\
&\equiv &\frac{1}{\sqrt{l!m!n!k!}}D_{1}^{\dagger l}D_{2}^{\dagger
m}D_{3}^{\dagger n}D_{4}^{\dagger k}\left\vert \mathbf{0}\right\rangle ,
\end{eqnarray}%
with the ground state $|\mathbf{0}\rangle \equiv |0,0\rangle _{ab}\otimes |%
\mathbf{b}\rangle $. Here, $|0,0\rangle _{ab}$ is the vacuum state of the
two quantized optical fields, $|\mathbf{b}\rangle =\otimes
\prod_{j}|b\rangle _{j}$ is the ground state for all atoms with the
definition $C|\mathbf{b}\rangle =A|\mathbf{b}\rangle =0$. The eigenvalue of
the state $\left\vert \Psi _{lmnk}\right\rangle $ is
\begin{equation}
\varepsilon _{lmnk} = (l-m)\varepsilon _{1}+(n-k)\varepsilon _{3}.
\end{equation}

It should be pointed out that the polaritons $D_{i}$ obtained in present
cyclic ensemble are different from the dark state polaritons in the $\Lambda
$-type ensemble~\cite{sun-prl}. The latter are the dark state polaritons and
commute with the interaction Hamiltonian, but the former ones do not commute
with the interaction Hamiltonian, and also are not dark state polaritons.

\section{Generation of quantum entanglements and the coherent output}

Now, we study how to generate the entangled states by using solutions of the
polaritons and their eigenstates. We first calculate the dynamical evolution
driven by the Hamiltonian~(\ref{h3}) with the constants $\Omega $, $g_N $
and $\varphi \equiv 0$. In this case, the polariton operators in Eqs.~(\ref%
{pola12}-\ref{pola34}) are simplified to
\begin{eqnarray}
D_{1,2} &=&\frac{\sin \theta }{\sqrt{2}}(a\pm b)+\frac{\cos \theta }{\sqrt{2}
}(C\pm A),\text{ }  \label{dd1} \\
D_{3,4} &=&\frac{\cos \theta }{\sqrt{2}}(a\pm b)-\frac{\sin \theta }{\sqrt{2}%
}(C\pm A).\text{ }  \label{dd2}
\end{eqnarray}%
The Heisenberg equations
\begin{equation*}
\partial _{t}D_{j}=-i[D_{j},H]=-i\varepsilon _{j}D_{j}\text{ (}j=1,2,3,4%
\text{)}
\end{equation*}%
describe the time evolution of the normal modes of the polaritons
\begin{equation}
D_{j}(t)=e^{-i\phi _{j}}D_{j}(0)\text{ (}j=1,2,3,4\text{)},  \label{Djt}
\end{equation}%
where $\phi _{j}\equiv \phi _{j}(t)=\varepsilon _{j}t$ is a time-dependant
phase. Since the physical modes can be expressed by the normal modes as
\begin{eqnarray}
a &=&\frac{1}{\sqrt{2}}\left[ (D_{1}+D_{2})\sin \theta +(D_{3}+D_{4})\cos
\theta \right] ,  \label{eq:22} \\
b &=&\frac{1}{\sqrt{2}}\left[ (D_{1}-D_{2})\sin \theta +(D_{3}-D_{4})\cos
\theta \right] ,  \label{eq:23} \\
A &=&\frac{1}{\sqrt{2}}\left[ (D_{1}-D_{2})\cos \theta -(D_{3}-D_{4})\sin
\theta \right] , \\
C &=&\frac{1}{\sqrt{2}}\left[ (D_{1}+D_{2})\cos \theta -(D_{3}+D_{4})\sin
\theta \right] ,
\end{eqnarray}%
the time-dependent operators $a(t)$, $b(t)$, $A(t)$, and $C(t)$ (also $%
a^{\dagger }(-t)$, $b^{\dagger }(-t)$, $A^{\dagger }(-t)$, and $C^{\dagger
}(-t)$) can be obtained by a straightforward replacement $D_{j}\rightarrow
D_{j}(0)\exp [-i\phi _{j}(t)]$ ($D_{j}^{\dagger }\rightarrow D_{j}^{\dagger
}(0)\exp [-i\phi _{j}(t)]$). The explicit expressions for these operators
are given in the appendix (see, Appendix~\ref{appendix}).

In what follows in this section, we will investigate the dynamical evolution
of the above cyclic system and show that the entanglement and the
information exchange between two optical modes can occur in the present
cyclic system for an initial direct-product Fock states of two optical
modes. We will also show that the atomic coherent excitation and coherent
output of photons can occur when the system is initially in a direct-product
coherent states of two optical modes.

\subsection{ Generation of entanglement between two optical modes}

If the system is initially in the two-mode photon number state
\begin{equation*}
\left\vert \psi (0)\right\rangle =\frac{1}{\sqrt{m!n!}}a^{\dagger
m}b^{\dagger n}\left\vert \mathbf{0}\right\rangle ,
\end{equation*}%
where $\left\vert \mathbf{0}\right\rangle \equiv \left\vert 0,0\right\rangle
_{ab}\otimes \left\vert \mathbf{b}\right\rangle \equiv \left\vert
0,0,0,0\right\rangle _{abAC}$ is the ground state of the system. Then,
according to Eqs.~(\ref{eq:22}-\ref{eq:23}), at time $t$, the wavefunction
can be expressed as
\begin{eqnarray}
\left\vert \psi (t)\right\rangle &=&\frac{1}{\sqrt{m!n!}}\left[ a^{\dagger
}(-t)\right] ^{m}\left[ b^{\dagger }(-t)\right] ^{n}\left\vert \mathbf{0}%
\right\rangle  \notag \\
&=&\frac{1}{\sqrt{m!n!}}[F_{a}^{a}(t)a^{\dagger }(0)+F_{b}^{a}(t)b^{\dagger
}(0)  \notag \\
&&+F_{A}^{a}(t)A^{\dagger }(0)+F_{C}^{a}(t)C^{\dagger }(0)]^{m}
\label{evolution} \\
&&\times \lbrack F_{a}^{b}(t)a^{\dagger }(0)+F_{b}^{b}(t)b^{\dagger }(0)
\notag \\
&&+F_{A}^{b}(t)A^{\dagger }(0)+F_{C}^{b}(t)C^{\dagger }(0)]^{n}\left\vert
\mathbf{0}\right\rangle ,  \notag
\end{eqnarray}%
with the time-dependent coefficients $F_{\beta }^{\alpha }(t)$ ($\alpha $, $%
\beta =a$, $b$,\ $A$, $C$) given in the Appendix.

Eq.~(\ref{evolution}) shows that the entanglement between the optical modes
and atomic collective modes can be generated when the coefficients in Eq.~(%
\ref{evolution}) satisfy certain conditions. However, in the following, we
will only focus on how to generate quantum entanglement between two optical
modes. When the coefficients of the atomic operators $A$ and $C$ of Eq.~(\ref%
{evolution}) vanish in some instants or in the certain limit, i.e.,
\begin{equation}
F_{A}^{a}(t)=F_{C}^{a}(t)=F_{A}^{b}(t)=F_{C}^{b}(t)=0,  \label{equ0}
\end{equation}%
the state $\vert \psi (t)\rangle$ in Eq.~(\ref{evolution}) only contains the
variables of photons, namely,
\begin{eqnarray}  \label{eq:28}
\left\vert \psi (t)\right\rangle &=&\frac{1}{\sqrt{m!n!}}[F_{a}^{a}(t)a^{%
\dagger }(0)+F_{b}^{a}(t)b^{\dagger }(0)]^{m} \\
&&\times \lbrack F_{a}^{b}(t)a^{\dagger }(0)+F_{b}^{b}(t)b^{\dagger
}(0)]^{n}\left\vert \mathbf{0}\right\rangle.  \notag
\end{eqnarray}

There are three cases in which the above photon state~(\ref{eq:28}) can be
generated during the dynamical evolution satisfying Eq. (\ref{equ0}).

\textit{Case I}: When the classical field is strongly coupled to the atomic
ensemble such that the coupling constants $\Omega $ and $g_{N}$ satisfy the
condition $\Omega /g_{N}\rightarrow +\infty $, then $\theta \approx 0$ and
the polariton operators in Eqs.~(\ref{pola12}-\ref{pola34}) can be
simplified to
\begin{equation}
D_{1,2}=\frac{1}{\sqrt{2}}(C\pm A),\text{ }D_{3,4}=\frac{1}{\sqrt{2}}(a\pm
b).  \label{eq:30}
\end{equation}%
In such condition, the time-dependent state in Eq.~(\ref{evolution}) becomes
into
\begin{eqnarray}
\left\vert \psi (t)\right\rangle &=&\frac{1}{\sqrt{m!n!}}[a^{\dagger
}(0)\cos \phi _{3}-ib^{\dagger }(0)\sin \phi _{3}]^{m}  \label{state-t6} \\
&&\times \lbrack -ia^{\dagger }(0)\sin \phi _{3}+b^{\dagger }(0)\cos \phi
_{3}]^{n}\left\vert \mathbf{0}\right\rangle .  \notag
\end{eqnarray}%
Eq.~(\ref{state-t6}) shows that the entanglement of optical modes $a$ and $b$
is obtained if the condition $\phi _{3}(t)\neq l\pi /2$ with the integer $l$
is satisfied. When $\phi _{3}(t)=\pi /2\ $(mod$\,\pi $), the state $%
\left\vert \psi (t)\right\rangle =a^{\dagger n}b^{\dagger m}\left\vert
\mathbf{0}\right\rangle /\sqrt{m!n!}$ with a negligibly global factor. This
process means that the information between the modes $a$ and $b$ is
exchanged. When $\phi _{2}(t)=0$ ($\text{mod}\,\pi $), the state $\left\vert
\psi (t)\right\rangle =\left\vert \psi (0)\right\rangle =$ $a^{\dagger
m}b^{\dagger n}\left\vert \mathbf{0}\right\rangle /\sqrt{m!n!}$, which
returns to the initial state.

\textit{Case II}: When the coupling of the classical field to the atomic
ensemble is much stronger than that of the quantized fields. That is, the
Rabi frequencies satisfy the condition $\Omega /g_{N}\rightarrow -\infty $,
then $\theta =\pi /2$. In this case, the polariton operators can be
simplified to
\begin{equation}
D_{1,2}=\frac{1}{\sqrt{2}}(a\pm b),\text{ }D_{3,4}=-\frac{1}{\sqrt{2}}(C\pm
A).  \label{eq:33}
\end{equation}%
The state in Eq. (\ref{evolution}) becomes into
\begin{eqnarray}
\left\vert \psi (t)\right\rangle &=&\frac{1}{\sqrt{m!n!}}[a^{\dagger
}(0)\cos \phi _{1}-ib^{\dagger }(0)\sin \phi _{1}]^{m} \\
&&\times \lbrack -ia^{\dagger }(0)\sin \phi _{1}+b^{\dagger }(0)\cos \phi
_{1}]^{n}\left\vert \mathbf{0}\right\rangle .  \notag
\end{eqnarray}%
Similar to the case I, the entanglement between optical modes $a$ and $b$
can also be obtained. When $\phi _{1}(t)=0$ ($\text{mod}$ $\pi $), the state
$\left\vert \psi (t)\right\rangle =\left\vert \psi (0)\right\rangle $; when $%
\phi _{1}(t)=\pi /2$ ($\text{mod}$ $\pi $), the state $|\psi (t)\rangle
=a^{\dagger n}b^{\dagger m}|\mathbf{0}\rangle /\sqrt{m!n!}$. Same as the
case I, the above processes mean that the state can return to the initial
one, or the quantum information between modes $a$ and $b$ can be exchanged
in some instants.

\textit{Case III}: Under the condition of $\phi _{1}(t)=\phi _{3}(t)+2\pi l $
with the integer $l$, we can also obtain the similar results as the above.
Comparing with the cases I and II that are in the special limit of the ratio
$|\Omega /g_{N}|$, here we consider a general case of $\Omega /g_{N}$. At
the instants $t_{s}=st_{0}$ ($s=0,1,2,...$) with $t_{0}=2\pi /\sqrt{\Omega
^{2}+4g_{N}^{2}}$, the time-dependant phases satisfy%
\begin{equation}
\phi _{1}(t_{s})=\phi _{3}(t_{s})\ (\text{mod}\text{ }2\pi ),
\end{equation}%
then Eq.~(\ref{evolution}) is
\begin{eqnarray}
\left\vert \psi (t_{s})\right\rangle &=&\frac{1}{\sqrt{m!n!}}[a^{\dagger
}(0)\cos \phi _{s}-ib^{\dagger }(0)\sin \phi _{s}]^{m} \\
&&\times \lbrack -ia^{\dagger }(0)\sin \phi _{s}+b^{\dagger }(0)\cos \phi
_{s}]^{n}\left\vert \mathbf{0}\right\rangle ,  \notag
\end{eqnarray}%
which is a two-mode photonic entangled state when $\phi_{s}\neq l\pi/2$ with
the integer $l$, where $\phi _{s}\equiv \phi (t_{s})=\phi _{1}(t_{s})=\phi
_{3}(t_{s})$ ($\text{mod}$ $\pi $).

Moreover, when the the special value of $\Omega /g_{N}$ is taken in case
III, the modes $a$ and $b$ can be disentangled in some certain instants. For
examples, if $\Omega ^{2}/g_{N}^{2}=4p^{2}/(q^{2}-p^{2})$ ($p$, $q$ are
integers), then at time $\tau _{s}^{(q)}=qt_{s}=2\pi qs/\sqrt{\Omega
^{2}+4g_{N}^{2}}\ $($s=0,1,2,...$), one has $\left\vert \cos \phi
_{1,3}(\tau _{s}^{(q)})\right\vert =1$, and
\begin{equation}
\left\vert \psi (\tau _{s}^{(q)})\right\rangle =\frac{1}{\sqrt{m!n!}}%
a^{\dagger m}(0)b^{\dagger n}(0)\left\vert \mathbf{0}\right\rangle ,
\end{equation}%
which is just the same state as the initial one. If $\Omega
^{2}/g_{N}^{2}=p^{2}/(q^{2}-p^{2})$ ($p$, $q$ are integers), then at time $%
\tau _{s}^{(q)}=qt_{s}=2\pi qs/\sqrt{\Omega ^{2}+4g_{N}^{2}}$ ($s=0,1,2,...$%
), one has $\left\vert \sin \phi _{1,3}(\tau _{s}^{(q)})\right\vert =1$, and
then we have
\begin{equation}
\left\vert \psi (\tau _{s}^{(q)})\right\rangle =\frac{1}{\sqrt{m!n!}}%
a^{\dagger n}(0)b^{\dagger m}(0)\left\vert \mathbf{0}\right\rangle ,
\end{equation}%
which means the information carried by two optical modes has been exchanged
between modes $a$ and $b$.

For the entangled states generated by the above three cases, we can further
calculate their entanglement degree in order to make them more clearly. In
fact, for each pure state, the entanglement can be defined as the entropy of
either of the two subsystems~\cite{bennet96,wooters,entangle}. For example,
the expression of entanglement for Eq.~(\ref{state-t6}) is given as
\begin{equation}
E(\left\vert \psi (t)\right\rangle )=-\mathrm{Tr}(\rho _{a}\log _{2}\rho
_{a}(t)),
\end{equation}%
where
\begin{equation}
\rho _{a}(t)=\mathrm{Tr}_{b}\rho (t)=\mathrm{Tr}_{b}(\left\vert \psi
(t)\right\rangle \left\langle \psi (t)\right\vert )
\end{equation}%
is the reduced density matrix of mode $a$. For simplicity, we consider the
case of $m=n=1$ in Eq.~(\ref{state-t6}). In this case, we have
\begin{eqnarray}
\rho _{a}(t) &=&\mathrm{Tr}_{b}(\left\vert \psi (t)\right\rangle
\left\langle \psi (t)\right\vert )  \notag \\
&=&2\sin ^{2}\phi _{3}\cos ^{2}\phi _{3}(|0\rangle _{aa}\langle 0|+|2\rangle
_{aa}\langle 2|) \\
&&+(\cos ^{2}\phi _{3}-\sin ^{2}\phi _{3})^{2}|1\rangle _{aa}\langle 1|,
\notag
\end{eqnarray}%
and the entanglement is given as%
\begin{eqnarray}
&&E(\left\vert \psi (t)\right\rangle )=-\mathrm{Tr}(\rho _{a}\log _{2}\rho
_{a}(t))  \notag \\
&=&(\cos ^{2}\phi _{3}-\sin ^{2}\phi _{3})^{2}\log _{2}(\cos ^{2}\phi
_{3}-\sin ^{2}\phi _{3})^{2}  \notag \\
&&+4\sin ^{2}\phi _{3}\cos ^{2}\phi _{3}\log _{2}(2\sin ^{2}\phi _{3}\cos
^{2}\phi _{3}),  \label{entanglement}
\end{eqnarray}%
here $\phi _{3}(t)\equiv \varepsilon _{3}t=(\Omega -\sqrt{\Omega
^{2}+4g_{N}^{2}})t/2$.

\begin{figure}[h]
\begin{center}
\includegraphics[width=8cm,height=4cm]{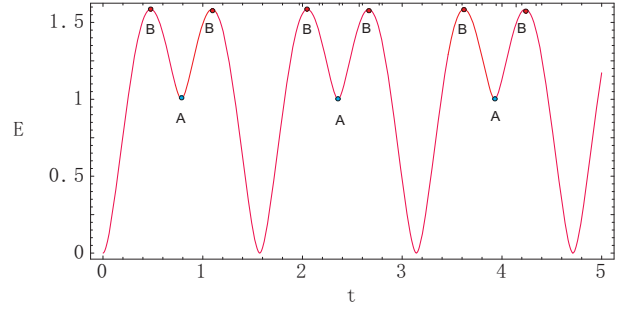}
\end{center}
\caption{(Color online) The entanglement in the state
Eq.~(\protect\ref{entanglement}) vs the time $t$ [in units of
$1/\protect\varepsilon _{3}$]. Points A show the maximally entangled
states for $2\times 2$-dimension. Points B show the maximally entangled
states for $3\times 3$-dimension.} \label{fig2}
\end{figure}

Figure~\ref{fig2} shows the entanglement in Eq.~(\ref{entanglement}) during
the time evolution. In fact, at time $t=(k\pm 1/4)\pi /\varepsilon _{3}$ ($%
k=0,1,2,...$), $\left\vert \psi (t)\right\rangle =(\left\vert
20\right\rangle _{ab}+\left\vert 02\right\rangle _{ab})/\sqrt{2}$ and $%
E(\left\vert \psi (t)\right\rangle )=1$, as shown in Fig.~\ref{fig2} with
those $A$ points. It means that $\left\vert \psi (t)\right\rangle $ is a
maximally entangled state for two optical modes in the two dimensional
space, constructed by the states $|2,0\rangle _{ab}$ and $|0,2\rangle _{ab}$
. At the time $t=[k\pi \pm \arcsin (\sqrt{(3+\sqrt{3})/6})]/\varepsilon _{3}$
or $t=[k\pi \pm \arcsin (\sqrt{(3-\sqrt{3})/6})]/\varepsilon _{3}$ ($%
k=0,1,2,... $), it has $\left\vert \psi (t)\right\rangle =(\left\vert
20\right\rangle _{ab}+\left\vert 02\right\rangle _{ab}\pm i\left\vert
11\right\rangle _{ab})/\sqrt{3}$ and $E(\left\vert \psi (t)\right\rangle
)=\log _{2}3$, which shows that $\left\vert \psi (t)\right\rangle $ is a
maximally entangled state in the three dimensional space, constructed by the
states $|20\rangle _{ab}$, $|02\rangle _{ab}$, and $|11\rangle _{ab}$ of two
optical modes (see the points B in Fig.~\ref{fig2}). These so-called
maximally entangled states are very useful in the field of quantum
information, e.g., to implement the quantum teleportation.

In this subsection, we have studied the entanglement and the information
exchange between two optical modes in the present cyclic system for an
initial direct-product Fock states of optical modes. Such an entanglement or
information exchange phenomenon cannot be occurred in a non-cyclic
three-level system, e.g. $V$-type three-level system~\cite{Jin03}.
Physically, the classical field, which is used to couple two higher levels,
assists to implement the entanglement (or information exchange) between
these optical modes. This can be seen from Eq.~(\ref{hh2}), in a non-cyclic
three-level $V$-type system obtained from the $\Delta$-type system with $%
\Omega =0$, two optical modes only interact independently with two
collective excitation modes respectively. So it can not realize the
entanglement between two optical modes.

\subsection{Coherent output of collective excitations and photons}

Here we study the dynamical evolution of the cyclic system when the system
is initially in the several kinds of coherent state as follows.

i) If the system is initially in a direct product state of coherent states
for four modes $a$, $b$, $A$ and $C$:
\begin{eqnarray*}
\left\vert \psi (0)\right\rangle  &=&D_{a}(\alpha )D_{b}(\beta )D_{A}(\zeta
)D_{C}(\eta )\left\vert \mathbf{0}\right\rangle  \\
&\equiv &\left\vert \alpha ,\beta ,\zeta ,\eta \right\rangle _{abAC},
\end{eqnarray*}%
where $D_{Q}(\gamma )=\exp [\gamma Q^{\dagger }-\text{h.c.}]$ ($Q=a,b,A,C$)
is the displacement operator. Then the state evolves into
\begin{equation}
\left\vert \psi (t)\right\rangle =D_{a}(\alpha ^{\prime })D_{b}(\beta
^{\prime })D_{A}(\zeta ^{\prime })D_{C}(\eta ^{\prime })\left\vert \mathbf{0}%
\right\rangle   \label{state-t9}
\end{equation}%
with
\begin{eqnarray*}
\alpha ^{\prime }(t) &=&\alpha F_{a}^{a}(t)+\beta F_{a}^{b}(t)+\zeta
F_{a}^{A}(t)+\eta F_{a}^{C}(t), \\
\beta ^{\prime }(t) &=&\alpha F_{b}^{a}(t)+\beta F_{b}^{b}(t)+\zeta
F_{b}^{A}(t)+\eta F_{b}^{C}(t), \\
\zeta ^{\prime }(t) &=&\alpha F_{A}^{a}(t)+\beta F_{A}^{b}(t)+\zeta
F_{A}^{A}(t)+\eta F_{A}^{C}(t), \\
\eta ^{\prime }(t) &=&\alpha F_{C}^{a}(t)+\beta F_{C}^{b}(t)+\zeta
F_{C}^{A}(t)+\eta F_{C}^{C}(t).
\end{eqnarray*}%
Here, the relation
\begin{eqnarray*}
&&U(t)D_{Q}(\gamma )U^{\dagger }(t) \\
&=&\exp [\gamma Q^{\dagger }(-t)-\text{h.c.}] \\
&=&\exp [\gamma (F_{a}^{Q}(t)a^{\dagger }(0)+F_{b}^{Q}(t)b^{\dagger }(0) \\
&&+F_{A}^{Q}(t)A^{\dagger }(0)+F_{C}^{Q}(t)C^{\dagger }(0))-\text{h.c.}]
\end{eqnarray*}%
has been used in Eq.~(\ref{state-t9}). Eq.~(\ref{state-t9}) shows that any
initial direct-product coherent state is still a direct-product coherent
state during the time evolution. However, the intensity of each mode varies
with the time evolution.

ii) If the atoms are initially in the ground states, but one of the optical
modes, e.g., mode $a$, is initially in a coherent state
\begin{equation}
\left\vert \psi (0)\right\rangle =D_{a}(\alpha )\left\vert \mathbf{0}%
\right\rangle =\left\vert \alpha ,0,0,0\right\rangle _{abAC}.
\label{state-t7}
\end{equation}%
Then, at time $t$, the state is
\begin{equation*}
\left\vert \psi (t)\right\rangle =\left\vert \alpha ^{\prime }(t),\beta
^{\prime }(t),\zeta ^{\prime }(t),\eta ^{\prime }(t)\right\rangle _{abAC},
\end{equation*}%
where%
\begin{eqnarray*}
\alpha ^{\prime }(t) &=&\alpha F_{a}^{a}(t),\text{ }\beta ^{\prime
}(t)=\alpha F_{b}^{a}(t), \\
\zeta ^{\prime }(t) &=&\alpha F_{A}^{a}(t),\text{ }\eta ^{\prime }(t)=\alpha
F_{C}^{a}(t).
\end{eqnarray*}%
This means that \emph{a new coherent optical field} of mode $b$ and two new
coherent atomic collective excitations are generated.

iii) If the atoms are initially in the ground states, and the two optical
modes are initially in their coherent states:
\begin{equation}
\left\vert \psi (0)\right\rangle =D_{a}(\alpha )D_{b}(\beta )\left\vert
\mathbf{0}\right\rangle =\left\vert \alpha ,\beta ,0,0\right\rangle _{abAC},
\end{equation}%
then, the evolved state will be direct-product coherent state of four modes:
\begin{equation}
\left\vert \psi (t)\right\rangle =D_{a}(\alpha ^{\prime })D_{b}(\beta
^{\prime })D_{A}(\zeta ^{\prime })D_{C}(\eta ^{\prime })\left\vert \mathbf{0}%
\right\rangle   \label{state-t11}
\end{equation}%
with
\begin{eqnarray*}
\alpha ^{\prime }(t) &=&\alpha F_{a}^{a}(t)+\beta F_{a}^{b}(t), \\
\beta ^{\prime }(t) &=&\alpha F_{b}^{a}(t)+\beta F_{b}^{b}(t), \\
\zeta ^{\prime }(t) &=&\alpha F_{A}^{a}(t)+\beta F_{A}^{b}(t), \\
\eta ^{\prime }(t) &=&\alpha F_{C}^{a}(t)+\beta F_{C}^{b}(t).
\end{eqnarray*}%
This means the initial optical modes can lead to the coherent output of new
modes of atomic collective excitation.

iv) If the two optical modes are initially in the vacuum state, but the two
atomic collective excitation modes are initially in the coherent states,
\begin{equation}
\left\vert \psi (0)\right\rangle =D_{A}(\zeta )D_{C}(\eta )\left\vert
\mathbf{0}\right\rangle .
\end{equation}%
Then, the state will evolve into a direct-product state of the two optical
modes and the atomic excitation modes:
\begin{equation}
\left\vert \psi (t)\right\rangle =D_{a}(\alpha ^{\prime })D_{b}(\beta
^{\prime })D_{A}(\zeta ^{\prime })D_{C}(\eta ^{\prime })\left\vert \mathbf{0}%
\right\rangle  \label{state-t12}
\end{equation}%
with%
\begin{eqnarray*}
\alpha ^{\prime }(t) &=&\zeta F_{a}^{A}(t)+\eta F_{a}^{C}(t), \\
\beta ^{\prime }(t) &=&\zeta F_{b}^{A}(t)+\eta F_{b}^{C}(t), \\
\zeta ^{\prime }(t) &=&\zeta F_{A}^{A}(t)+\eta F_{A}^{C}(t), \\
\eta ^{\prime }(t) &=&\zeta F_{C}^{A}(t)+\eta F_{C}^{C}(t).
\end{eqnarray*}

Eqs.~(\ref{state-t11}) and (\ref{state-t12}) show that the coherent optical
modes or coherent atomic excitation modes will result in the generation of
the coherent atomic excitation modes or coherent optical modes in the cyclic
atomic ensemble system.

v) We now consider that one of the optical modes, e.g., the mode $a$, is
initially in odd or even coherent states $\mathcal{N}(\left\vert \alpha
\right\rangle _{a}\pm \left\vert -\alpha \right\rangle _{a})$ with the
normalization constant $\mathcal{N}=(2\pm 2e^{-2|\alpha |^{2}})^{-1/2}$. But
another optical mode $b$ is in the vacuum state, and also the two atomic
modes are in their ground states. That is, the system is initially in the
state
\begin{equation}
\left\vert \psi (0)\right\rangle =\mathcal{N}(\left\vert \alpha
\right\rangle _{a}\pm \left\vert -\alpha \right\rangle
_{a})\otimes \left\vert 0,0,0\right\rangle _{bAC}.  \label{eq:49}
\end{equation}%
At instant $\tau $ when $\phi _{1}(\tau )=\phi _{3}(\tau )=\phi (\tau )$
(mod $2\pi $), the state will evolve to a so-called \textit{entangled
coherent state} of these two optical modes~\cite%
{Agarwal89,Sanders92,Solano03}
\begin{eqnarray}
|\psi (\tau )\rangle &=&|\mathbf{0}\rangle _{AC}\otimes \mathcal{N}[|\alpha
\cos \phi ,-i\alpha \sin \phi \rangle _{ab}  \notag \\
&&\pm |-\alpha \cos \phi ,i\alpha \sin \phi \rangle _{ab}].  \label{state-t5}
\end{eqnarray}%
When $\phi (\tau )$\ in Eq.~(\ref{state-t5}) satisfies $\phi (\tau )=\pi /4$
(mod $\pi $), the state will be in a maximally entangled state~\cite{vanEnk}%
. These states have recently been proposed as an important tool in the
theories and experiments relating to the quantum information processing~\cite%
{Munro00,Wang01,Jeong01}.

Specially, if $\phi (\tau )=0$ (mod $\pi $) at certain time $\tau $, the
instantaneous state in Eq.~(\ref{state-t5}) returns to the initial state in
Eq.~(\ref{eq:49}). If $\phi (\tau )=\pi /2$ (mod $\pi $) at certain time $%
\tau $, the instantaneous state in Eq.~(\ref{state-t5}) becomes into
\begin{equation}
\left\vert \psi (\tau )\right\rangle =\mathcal{N}(\left\vert
\alpha \right\rangle _{b}\pm \left\vert -\alpha \right\rangle
_{b})\otimes \left\vert 0,0,0\right\rangle_{aAC}. \label{state-t3}
\end{equation}%
It means that a new coherent state for the mode $b$ is generated. It also
shows that the quantum information is transferred from mode $a$ to mode $b$.

So far we have showed the atomic coherent excitation (or coherent
output of photons) when the optical fields (or atomic
collective-excitations) are initially in the coherent states.
Moreover, if one of the optical modes is initially in an odd (or
even) coherent state but another one is in the vacuum state, the
system will evolve to an entangled state for two optical modes
with coherent states each. When a special condition is satisfied,
the information can be transferred from the first optical mode to
the second one, which has been described in Eqs.~(\ref{eq:49}) and
(\ref{state-t3}). These interesting results is due to the
classical optical field, which induces the atomic transition
between two higher states. However these interesting phenomena can
not be found in the non-cyclic three-level $V$-type atomic
ensemble, where only the classical optical field is removed as a
comparison with the cyclic system.

\section{The state storage of photons based on adiabatic manipulation}

In section III, we have studied the dynamic properties of the atomic
ensemble with the cyclic transitions. Here, we consider the adiabatic
evolution of the cyclic atomic ensemble, controlled by the time-dependent
classical field. In this case, the Rabi frequency $\Omega$ should become
into the time-dependent one, i.e., $\Omega (t)$. Using the diagonalized
Hamiltonian~(\ref{hh2}) and following the method of collective excitations
shown in Ref.~\cite{sun-prl}, we will discuss how to transfer the
information from the two quantized light fields to the atomic ensemble by
the adiabatic passage.

In the following, the value of $\varphi $ is fixed, i.e., $\varphi =0$, but $%
\Omega $ will be changed within the range ($-\infty ,+\infty $) according to
the constant $g_{N}$. In general, the polariton operators $D_{i}$ $%
(i=1,\,2,\,3,\,4)$ consist of two photonic modes and two atomic
collective excitation modes. For simplicity, we consider two
simple cases for $D_{i}$. One is $\Omega /g_{N}\rightarrow +\infty
$. In this limit, $\theta \rightarrow 0$, and the polariton
operators are given in Eqs.~(\ref{eq:30}) with the relative values
$\varepsilon _{1}\equiv -\varepsilon _{2}\rightarrow \Omega $,
$\varepsilon _{3}\equiv -\varepsilon _{4}\rightarrow 0$. Another
is $\Omega /g_{N}\rightarrow -\infty $, then we have $\theta
\rightarrow \pi /2$, and the polariton operators become into
Eq.~(\ref{eq:33}) with the relative values $\varepsilon _{1}\equiv
-\varepsilon _{2}\rightarrow 0$, $\varepsilon _{3}\equiv
-\varepsilon _{4}\rightarrow \Omega $.

The analysis on Eq.~(\ref{eq:16}) shows that when $\Omega $ varies in the
range ($-\infty ,+\infty $), $\theta $ will vary in the range ($0,\pi /2$).
In the above two limit cases, the polariton operators $D_{1,2}$ (or $D_{3,4}$%
) consist of only the optical modes $a,\,b$ (or only the atomic collective
excitation modes $A,\,C$). This implies that the information can be
transferred from two quantized light fields to the atomic ensemble, and then
can also be stored in the atomic ensemble, as given in Ref.~\cite{sun-prl},

For example, initially, if we set $\Omega (t)/g_{N}\rightarrow +\infty $,
and the system is in a direct-product Fock state of two optical modes
\begin{equation}
\left\vert \Psi (0)\right\rangle =\left\vert m,n\right\rangle _{ab}\otimes
\left\vert \mathbf{b}\right\rangle =\frac{a^{\dagger m}b^{\dagger n}}{\sqrt{%
m!n!}}\left\vert \mathbf{0}\right\rangle .  \label{eq:52}
\end{equation}%
Then, using expressions of the polariton operators in Eq.~(\ref{eq:30}),
Eq.~(\ref{eq:52}) can be rewritten as a superposition
\begin{equation}
\left\vert \Psi (0)\right\rangle =\sum_{j,k}f_{mn}^{jk}\left( D_{3}^{\dagger
}\right) ^{j+k}\left( D_{4}^{\dagger }\right) ^{m+n-j-k}\left\vert \mathbf{0}%
\right\rangle ,
\end{equation}%
where%
\begin{eqnarray*}
f_{mn}^{jk} &=&(-1)^{n-k}\frac{C_{m}^{j}C_{m}^{m-j}C_{n}^{k}C_{n}^{n-k}} {%
\sqrt{2^{m+n}m!n!}}, \\
C_{m}^{j} &=&\frac{m!}{j!(m-j)!}.
\end{eqnarray*}%
In the process of the adiabatical evolution, the state at time $t$ will be
\begin{equation}
\left\vert \Psi (t)\right\rangle
=\sum_{j,k}f_{mn}^{jk}U_{mn}^{jk}(t)[D_{3}^{\dagger
}(t)]^{j+k}[D_{4}^{\dagger }(t)]^{m+n-j-k}\left\vert \mathbf{0}\right\rangle
,  \label{state-t1}
\end{equation}%
where $U_{mn}^{jk}(t)$ is the relative dynamic phase:%
\begin{equation}
U_{mn}^{jk}(t)=(2j+2k-m-n)\exp \left[-i\int_{0}^{t}\varepsilon
_{3}(t^{\prime })\mathrm{d}t^{\prime }\right].
\end{equation}

When $\Omega (t)$ is adiabatically changed to $-\infty $ in certain time $%
\tau $, the state will be
\begin{eqnarray}
\left\vert \Psi (\tau )\right\rangle
&=&\sum_{j,k}f_{mn}^{jk}U_{mn}^{jk}(\tau )[D_{3}^{\dagger }(\tau
)]^{j+k}[D_{4}^{\dagger }(\tau )]^{m+n-j-k}\left\vert \mathbf{0}\right\rangle
\notag \\
&=&\sum_{j,k}f_{mn}^{jk}U_{mn}^{jk}(\tau )\left[ -\frac{A^{\dagger
}(0)+C^{\dagger }(0)}{\sqrt{2}}\right] ^{j+k}  \label{state-t} \\
&&\times \left[ \frac{A^{\dagger }(0)-C^{\dagger }(0)}{\sqrt{2}}\right]
^{m+n-j-k}\left\vert \mathbf{0}\right\rangle .  \notag
\end{eqnarray}%
Eq.~(\ref{state-t}) shows that the quantum information carried by optical
modes ($a$ and $b$) has been completely transferred to the atomic collective
excitation modes $A$ and $C$. Since atoms are local and robust, the above
adiabatic process means that the information of two quantized light fields
has been stored in an atomic ensemble.

Especially, for certain evolution path of $\Omega (t)$, if the relative
dynamic phase $U_{mn}^{jk}(\tau )\equiv 1$ holds for any integer $m,n,j,k$,
that is
\begin{equation}
\int_{0}^{\tau }\varepsilon _{3}(t)\mathrm{d}t=2l\pi
\end{equation}
with an integer $l$, $\left\vert \Psi (\tau )\right\rangle $ will have a
simple form
\begin{equation}
\left\vert \Psi (\tau )\right\rangle =(-1)^{m+n}\left\vert 0\right\rangle
_{a}\left\vert 0\right\rangle _{b}\left\vert n\right\rangle _{A}\left\vert
m\right\rangle _{C}.
\end{equation}%
That is, when adiabatically changing $\Omega /g_{N}:+\infty \rightarrow
-\infty $, under the special case of $U_{mn}^{jk}(\tau )\equiv 1$ for any $%
j,k,m$ and $n$, the state transfer can be realized as follows:
\begin{eqnarray}
&&\left\vert \Psi (0)\right\rangle =\left\vert m\right\rangle _{a}\left\vert
n\right\rangle _{b}\left\vert 0\right\rangle _{A}\left\vert 0\right\rangle
_{C}  \notag \\
&\rightarrow &\left\vert \Psi (\tau )\right\rangle =(-1)^{m+n}\left\vert
0\right\rangle _{a}\left\vert 0\right\rangle _{b}\left\vert n\right\rangle
_{A}\left\vert m\right\rangle _{C}.
\end{eqnarray}%
Such an adiabatic passage means $a\rightarrow -C$ and $b\rightarrow -A$, so
the initial state involved only for the optical modes will evolve to the
final state involved only for the atomic excitation modes. This also means
that the quantum information carried by the optical fields has been
transferred to and stored in the atomic ensemble.

An inverse adiabatic passage, which makes $\Omega (\tau )/g_{N}=-\infty $ $%
\rightarrow $ $\Omega (T)/g_{N}=+\infty $, will result in information
transfer from the atomic ensemble to the two optical modes, i.e., $%
C\rightarrow -a$ and $A\rightarrow -b$. And then, the initial atomic state $%
|n\rangle_{A}|m\rangle_{C}$ will evolve to the final state $\left\vert \Psi
(T)\right\rangle =\left\vert m\right\rangle _{a}\left\vert n\right\rangle
_{b}\left\vert 0\right\rangle _{A}\left\vert 0\right\rangle _{C}$ by the
inverse adiabatic passage with the relative dynamic phase $%
U_{mn}^{jk}(T)\equiv 1$.

It is \emph{worth stressing} that if the initial state is $\left\vert \Psi
(0)\right\rangle =\left\vert 0\right\rangle _{a}\left\vert 0\right\rangle
_{b}\left\vert m\right\rangle _{A}\left\vert n\right\rangle _{C}$, after an
adiabatically changing $\Omega /g_{N}:+\infty \rightarrow -\infty $, this
state will become into the Fock state of the two optical modes $a$ and $b$
(depicted according to the polaritons $D_{1,2}$), where we also assume that
the dynamic phase factor is $1$ during the adiabatic evolution, that is%
\begin{equation}
U_{mn}^{\prime jk}(\tau )\equiv 1\text{ (for any integer }m,n,j,k\text{),}
\end{equation}%
which is equal to $\int_{0}^{\tau }\varepsilon _{1}(t)\mathrm{d}t=2l\pi $
with an integer $l$.

The inverse adiabatical passage $\Omega /g_{N}:-\infty \rightarrow +\infty $
will result in the information carried by the optical fields to be
transferred to that by atomic ensemble.

So far, we have achieved the quantum information exchange between optical
fields and atomic ensemble with initially in the Fock states. For general
states, e.g.,
\begin{equation}
\left\vert \Psi (0)\right\rangle =\sum_{m,n}u_{mn}\left\vert m\right\rangle
_{a}\left\vert n\right\rangle _{b}\left\vert 0\right\rangle _{A}\left\vert
0\right\rangle _{C}\text{ }
\end{equation}%
or $\left\vert \Psi (0)\right\rangle =\left\vert \alpha \right\rangle
_{a}\left\vert \beta \right\rangle _{b}\left\vert 0\right\rangle
_{A}\left\vert 0\right\rangle _{C},$the information can also be transferred
in the similar way as done in Ref.~\cite{sun-prl}.

In Section III, we mainly study the generation of entanglement states
between two optical modes, and the quantum information transfer from one
optical mode to another optical mode by virtue of \textit{the dynamical
evolution}. In this section, we discuss the information transfer and storage
from the optical fields to the cyclic atomic ensemble through \textit{the
adiabatic passage}. Moreover, the quantum information can also be retrieved
from the atomic collective excitation modes. It is well known that photons
are non-local and not easy to be stored, but atoms are local and robust. The
above process provides a way to implement retrievable storage of the optical
information in an atomic ensemble.

\section{Conclusion}

We have investigated various protocols of quantum information processing
based on the photonic properties of the emission and excitation of a $\Delta
$-type (or cyclic) \textquotedblleft atomic" ensemble, which coupled to two
quantum optical fields and one classical field. The classical field controls
the coupling between two upper energy levels. By means of collective
operator approach, we studied the dynamical evolution and adiabatic
manipulation for such a novel system. Our results show that the two-mode
photon entanglement and quantum information exchange between two optical
modes can be realized when the optical modes are initially in the
direct-product Fock states or coherent states.

It is remarked that, even without symmetry broken, a three-level system can
also form a cyclic one. The electric-dipole interaction of the classic field
coupled to the two higher states in our model can be replaced by the
magnetic-dipole transition. However this magnetic-dipole interaction is
generally very weak compared with the electric-dipole interaction and
disposed as perturbation. The significant phenomenon\ of cyclic three-level
configuration only occurs in the systems with symmetry broken as given in
the present work.

We also need to point out that, it is the classical field to result in
various new phenomena, found in this paper. Without this classical field, it
would be impossible to generate the entangled states of the two optical
modes. As a quantum memory, the collective excitations of the $\Delta $-type
atomic ensemble can store the quantum information carried by two quantum
optical modes through the adiabatical manipulation. The corresponding
adiabatical evolution is realized by choosing certain classical Rabi
frequency $\Omega (t)$. We expect that our proposal can be confirmed and
implemented experimentally in the near future.

In this paper, we just consider an ideal case. Actually, in a
realistic system, two kinds of decoherence mechanisms may play a
role. The first one comes from the multi-mode radiation, which
will result in the collective decay of atoms from the excited
states to the ground state. This kind of radiation properties due
to atomic decay in present system will be investigated in a
following paper. The second decoherence effect is due to the
inhomogeneous coupling of atoms to the light fields. The influence
of inhomogeneous coupling has been studied in detail in a
two-level-atom ensemble by Sun \textit{et}
\textit{al.}~\cite{sunandyou}. In the realistic experiments, the
atoms can be fixed and the coupling can be taken as the constant
by a special design to avoid the decoherence of inhomogeneous
coupling. Our assumption for homogeneous coupling is reasonable in
a short interaction time.

\section{acknowledgments}

This work is supported by the NSFC (with grant Nos. 90203018, 10474104,
10447133, 10547106, 10574133, and 60433050) and the National Fundamental
Research Program of China (with Nos. 2001CB309310 and 2005CB724508). Y.
L. is also supported by the China Postdoctoral Science Foundation (with
No. 2004036309) and Hong Kong K. C. Wong Education Foundation.

\appendix

\section{Derivation of polariton operators}

\label{appendix}

Here we rewrite the forms of $D_{1,2,3,4}$ in Eqs. (\ref{dd1},\ref{dd2}):

\begin{eqnarray}
D_{1} &=&\frac{\sin \theta }{\sqrt{2}}(a+b)+\frac{\cos \theta }{\sqrt{2}}%
(A+C),\text{ }  \label{ddd1} \\
D_{2} &=&\frac{\sin \theta }{\sqrt{2}}(a-b)-\frac{\cos \theta }{\sqrt{2}}%
(A-C),\text{ } \\
D_{3} &=&\frac{\cos \theta }{\sqrt{2}}(a+b)-\frac{\sin \theta }{\sqrt{2}}%
(A+C), \\
D_{4} &=&\frac{\cos \theta }{\sqrt{2}}(a-b)+\frac{\sin \theta }{\sqrt{2}}%
(A-C),\text{ }  \label{ddd4}
\end{eqnarray}%
According to Eq. (\ref{h3}), it has%
\begin{equation*}
\partial _{t}D_{j}=-i[D_{j},H]=-i\varepsilon _{j}D_{j}\text{ (}j=1,2,3,4%
\text{)},
\end{equation*}%
and then%
\begin{equation}
D_{j}(t)\equiv e^{-i\phi _{j}}D_{j}(0)\text{ (}j=1,2,3,4\text{)},
\label{Djt2}
\end{equation}%
where $\phi _{j}(t)\equiv \varepsilon _{j}t$. Following Eqs. (\ref{ddd1})-(%
\ref{ddd4}), we can obtain the inverse transformation:%
\begin{eqnarray}
a &=&\frac{1}{\sqrt{2}}\left[ (D_{1}+D_{2})\sin \theta +(D_{3}+D_{4})\cos
\theta \right] ,\text{ }  \label{a} \\
b &=&\frac{1}{\sqrt{2}}\left[ (D_{1}-D_{2})\sin \theta +(D_{3}-D_{4})\cos
\theta \right] ,  \label{b} \\
A &=&\frac{1}{\sqrt{2}}\left[ (D_{1}-D_{2})\cos \theta -(D_{3}-D_{4})\sin
\theta \right] ,  \label{A} \\
C &=&\frac{1}{\sqrt{2}}\left[ (D_{1}+D_{2})\cos \theta -(D_{2}+D_{3})\sin
\theta \right] .  \label{C}
\end{eqnarray}%
By means of Eq. (\ref{Djt2}), we have%
\begin{eqnarray}
a(t) &=&\frac{1}{\sqrt{2}}[(D_{1}(0)e^{-i\phi _{1}}+D_{2}(0)e^{i\phi
_{1}})\sin \theta \\
&&+(D_{3}(0)e^{-i\phi _{3}}+D_{4}(0)e^{i\phi _{3}})\cos \theta ]  \notag \\
&\equiv &F_{a}^{a}(t)a(0)+F_{b}^{a}(t)b(0)  \notag \\
&&+F_{A}^{a}(t)A(0)+F_{C}^{a}(t)C(0),  \notag
\end{eqnarray}%
\begin{eqnarray}
b(t) &=&\frac{1}{\sqrt{2}}[(D_{1}(0)e^{-i\phi _{1}}-D_{2}(0)e^{i\phi
_{1}})\sin \theta \\
&&+(D_{3}(0)e^{-i\phi _{3}}-D_{4}(0)e^{i\phi _{3}})\cos \theta ]  \notag \\
&\equiv &F_{a}^{b}(t)a(0)+F_{b}^{b}(t)b(0)  \notag \\
&&+F_{A}^{b}(t)A(0)+F_{C}^{b}(t)C(0),  \notag
\end{eqnarray}%
\begin{eqnarray}
A(t) &=&\frac{1}{\sqrt{2}}[(D_{1}(0)e^{-i\phi _{1}}-D_{2}(0)e^{i\phi
_{1}})\cos \theta \\
&&-(D_{3}(0)e^{-i\phi _{3}}-D_{4}(0)e^{i\phi _{3}})\sin \theta ]  \notag \\
&\equiv &F_{a}^{A}(t)a(0)+F_{b}^{A}(t)b(0)  \notag \\
&&+F_{A}^{A}(t)A(0)+F_{C}^{A}(t)C(0),  \notag
\end{eqnarray}%
\begin{eqnarray}
C(t) &=&\frac{1}{\sqrt{2}}[(D_{1}(0)e^{-i\phi _{1}}+D_{2}(0)e^{i\phi
_{1}})\cos \theta \\
&&-(D_{3}(0)e^{-i\phi _{3}}+D_{4}(0)e^{i\phi _{3}})\sin \theta ]  \notag \\
&\equiv &F_{a}^{C}(t)a(0)+F_{b}^{C}(t)b(0)  \notag \\
&&+F_{A}^{C}(t)A(0)+F_{C}^{C}(t)C(0),  \notag
\end{eqnarray}%
where we have used $\phi _{1}(t)=-\phi _{2}(t)$ and $\phi _{3}(t)=-\phi
_{4}(t)\,$and the related coefficients are%
\begin{eqnarray*}
F_{a}^{a}(t) &=&F_{b}^{b}(t)=\cos \phi _{1}\sin ^{2}\theta +\cos \phi
_{3}\cos ^{2}\theta , \\
F_{b}^{a}(t) &=&F_{a}^{b}(t)=-i(\sin \phi _{1}\sin ^{2}\theta +\sin \phi
_{3}\cos ^{2}\theta ), \\
F_{A}^{a}(t) &=&F_{a}^{A}(t)=-i\sin \theta \cos \theta (\sin \phi _{1}-\sin
\phi _{3}), \\
F_{C}^{a}(t) &=&F_{a}^{C}(t)=\sin \theta \cos \theta (\cos \phi _{1}-\cos
\phi _{3}),
\end{eqnarray*}%
\begin{eqnarray*}
F_{A}^{b}(t) &=&F_{b}^{A}(t)=(\cos \phi _{1}-\cos \phi _{3})\sin \theta \cos
\theta , \\
F_{C}^{b}(t) &=&F_{b}^{C}(t)=-i\sin \theta \cos \theta (\sin \phi _{1}-\sin
\phi _{3}), \\
F_{A}^{A}(t) &=&F_{C}^{C}(t)=\cos \phi _{1}\cos ^{2}\theta +\cos \phi
_{3}\sin ^{2}\theta , \\
F_{C}^{A}(t) &=&F_{A}^{C}(t)=-i(\sin \phi _{1}\cos ^{2}\theta +\sin \phi
_{3}\sin ^{2}\theta ).
\end{eqnarray*}%
It also has%
\begin{eqnarray}
a^{\dagger }(-t) &\equiv &F_{a}^{a}(t)a^{\dagger }(0)+F_{b}^{a}(t)b^{\dagger
}(0)  \notag \\
&&+F_{A}^{a}(t)A^{\dagger }(0)+F_{C}^{a}(t)C^{\dagger }(0),\text{ }  \notag
\\
b^{\dagger }(-t) &\equiv &F_{a}^{b}(t)a^{\dagger }(0)+F_{b}^{b}(t)b^{\dagger
}(0)  \notag \\
&&+F_{A}^{b}(t)A^{\dagger }(0)+F_{C}^{b}(t)C^{\dagger }(0),  \notag \\
A^{\dagger }(-t) &\equiv &F_{a}^{A}(t)a^{\dagger }(0)+F_{b}^{A}(t)b^{\dagger
}(0)  \label{evolutionabAC} \\
&&+F_{A}^{A}(t)A^{\dagger }(0)+F_{C}^{A}(t)C^{\dagger }(0),  \notag \\
C^{\dagger }(-t) &\equiv &F_{a}^{C}(t)a^{\dagger }(0)+F_{b}^{C}(t)b^{\dagger
}(0)  \notag \\
&&+F_{A}^{C}(t)A^{\dagger }(0)+F_{C}^{C}(t)C^{\dagger }(0).  \notag
\end{eqnarray}%
%
%
%
%
%
%

\end{document}